\documentclass[aps,prb,twocolumn,groupedaddress,amssymb,amsmath]{revtex4}

\usepackage{graphics,graphicx}

\begin{document}

\title{Competing mechanisms for step meandering in unstable growth}

\author{Jouni Kallunki}
\email[]{jkallunk@theo-phys.uni-essen.de}
\affiliation{Fachbereich Physik, Universit\"at Essen, 45117 Essen, Germany}

\author{Joachim Krug}
\affiliation{Fachbereich Physik, Universit\"at Essen, 45117 Essen, Germany}

\author{Miroslav Kotrla}
\affiliation{Institute of Physics, Academy of Sciences of the Czech Republic, Na
Slovance 2, 182 21 Prague 8, Czech Republic}

\date{\today}

\begin{abstract}
The meander instability of a vicinal surface growing under step flow
conditions is studied within a solid-on-solid model. In the absence of 
edge diffusion the selected meander wavelength agrees quantitatively with
the continuum linear stability analysis of Bales and Zangwill [Phys. Rev. B 
{\bf 41}, 4400 (1990)]. In the presence of edge diffusion a local instability
mechanism related to kink rounding
barriers dominates, and the meander 
wavelength is set by one-dimensional nucleation. The long-time behavior of the
meander amplitude differs in the two cases, and
disagrees with the predictions of a nonlinear step
evolution equation [O. Pierre-Louis et al., Phys. Rev. Lett. {\bf 80}, 4221
(1998)]. The variation of the meander wavelength
with the deposition flux and 
with the activation barriers for step adatom detachment and step
crossing (the Ehrlich-Schwoebel barrier) is studied in detail. 
The interpretation of recent experiments on surfaces vicinal to 
Cu(100) [T. Maroutian et al., Phys. Rev. B {\bf 64}, 165401 (2001)] in 
the light of our results yields an estimate for the
kink rounding barrier at the close packed steps.     
\end{abstract}


\maketitle

\section{Introduction}
It has been shown in several experiments and computer simulations that 
during epitaxial growth on a vicinal crystal surface,
straight steps are unstable against the formation 
of an in-phase meander \cite{Exps,Rost96}.
The phenomenon was first predicted theoretically by Bales and Zangwill
(BZ) within a continuum theory \cite{Bales90}. According to BZ,
step meandering is caused by an energy barrier, the 
Ehrlich-Schwoebel (ES) barrier \cite{ES66}, 
which suppresses the attachment of surface atoms to the step 
from the terrace above. The preferential attachment from below
implies that protrusions in the step are amplified, leading
to a linear diffusional instability. 

Recent experimental measurements of the meander wavelength on vicinal
copper surfaces \cite{Maroutian99,Maroutian01} show significant
disagreement with the predictions of the BZ theory. 
This has lead to a search for alternative sources of 
instability \cite{Rusanen01,Rusanen02,Pierre-Louis01}. 
The most prominent alternative mechanism is the kink 
Ehrlich-Schwoebel effect (KESE), which invokes a kinetic barrier that
prevents atoms diffusing along step edges to cross corners or kinks
\cite{Pierre-Louis99,Ramana99,Politi00a}. In close analogy to 
the ES instability of singular crystal 
surfaces \cite{Politi96,Krug97,Politi00b}, this induces a destabilizing
mass current along the step.

In this article we present kinetic Monte Carlo (KMC) simulations of 
step meandering which display both types of 
instability within a single model. The relative importance of the KESE
vs. the BZ instability can be tuned through the facility of step edge diffusion.
By explicitly relating the parameters of the KMC
model to those of the continuum theory, we show that
the meander wavelength can be quantitatively predicted in both 
instability regimes. Our study thus proves the feasibility of
extracting kinetic barriers from experimental meander morphologies. 
The simulations also provide some insight into
the long-time behavior of the meander amplitude which can be compared
to the predictions of nonlinear continuum equations 
\cite{Pierre-Louis98b,Kallunki00,Gillet00}.

The model employed in our work is described in the next section.
Section \ref{Predictions} summarizes the predictions of continuum
theory for the meander wavelength, and explains how the 
material parameters of the continuum description are determined
for the KMC model. The simulation results are presented in Section
\ref{Results}. We provide some discussion
of the applicability of the KESE scenario to the experiments 
\cite{Maroutian01} on Cu(100) in Section \ref{Exps},
and conclude in Section \ref{Conclusions}.

\section{Model}
We employ a standard solid-on-solid model \cite{Kotrla96}, in which 
the crystal surface is represented by columns of integer 
height $h_{\bf r}$  on a square lattice of substrate sites ${\bf r}$.
The elementary processes are the deposition of atoms at 
rate $F$ and the hopping of adatoms to nearest neighbor sites
with a rate 
\begin{equation}
\label{rates0}
r=r_0 \exp(-E_a/k_B T). 
\end{equation}
The activation barrier 
$E_a$ depends on the local configuration through 
\begin{eqnarray}
 \label{rates1}
  E_a &=& E_{\mathrm S}+n_iE_{\mathrm n}+(n_i-n_f)\Theta(n_i-n_f)E_{\mathrm BB} \\
  &+& (m_i-m_f)\Theta(m_i-m_f)E_{\mathrm ES}, \nonumber
\end{eqnarray}
where $E_{\mathrm S}$ is the 
energy barrier for diffusion on a flat terrace, 
$E_{\mathrm n}$ is the contribution of a nearest neighbor bond
to the energy barrier,  
$E_{\mathrm BB}$ is an additional energy cost for bond breaking
and $E_{\mathrm ES}$ is the ES barrier; $n_i$ denotes the number 
of in-plane nearest neighbors before the hop and
$n_f$ after the hop, while $m_i, m_f$ 
are the number of next-nearest 
neighbors in the planes beneath and above 
before ($m_i$) and after ($m_f$) the hop. The Heaviside function
$\Theta(x) = 1$ if $x>0$ and 0 otherwise.

The implementation of the ES barrier through the change in the 
number of out-of-plane next nearest neighbors has been used
in several earlier growth studies \cite{Rost96,Smilauer95,Siegert96},
and the additional bond breaking energy $E_{\mathrm BB}$ was introduced
in the context of ion sputtering \cite{Smilauer93}. 
The rates defined by (\ref{rates0}) and (\ref{rates1})
satisfy detailed balance with respect to the Hamiltonian 
\begin{eqnarray}
 \label{Hamiltonian}
  H &=& \sum_{\langle {\bf r},{\bf r'} \rangle} 
[ E_{\mathrm K} |h_{\bf r}-h_{\bf r'}|  \\  
  &+& E_{\mathrm ES} (|h_{\bf r}-h_{\bf r'}|-1) 
  \Theta(|h_{\bf r} -h_{\bf r'}|-1)].
  \nonumber 
\end{eqnarray}
The sum runs over all nearest neighbor pairs, and
\begin{equation}
\label{EK}
E_{\mathrm K}=\frac{1}{2}(E_{\mathrm n}+E_{\mathrm BB})
\end{equation} 
is the energy per unit length of a single height 
step running along one of the
lattice axes; for the SOS model, this is also the
kink energy. The detailed balance condition is easily checked
by noting that the rates can be written as a product of Arrhenius 
(term proportional to $E_{\mathrm n}$ in (\ref{rates1})) 
and Metropolis 
(terms proportional to $E_{\mathrm{BB}}$ and 
$E_{\mathrm{ES}}$) 
dynamics \cite{Siegert96}, each of which fulfil detailed
balance with respect to part of the Hamiltonian (\ref{Hamiltonian}).

Setting $E_{\mathrm BB}=0$ the model, called hereafter \emph{model I},
does not include diffusion along the step edges, because 
the hopping rate along the step is equal to 
the rate of detachment from the step. 
Edge diffusion is facilitated compared to detachment
if $E_{\mathrm BB} > 0$ (called hereafter \emph{model II}). 
Model II also contains a kink ES barrier, since
atoms cannot round corners without detaching from the edge.

The simulations were carried out on rectangular lattices with
periodic boundary conditions 
in the step direction and helical boundary conditions in
the direction of the vicinality. The initial step spacing
was typically $l = 6$ (exceptions are noted in the 
figure captions). For both models
the values for the activation barriers were set to
$E_{\mathrm S} = 0.35$ eV, $E_{\mathrm n} = 0.21$ eV and $E_{\mathrm ES} = 0.15$ eV,
$E_{\mathrm BB} =0$ for model I  and for model II we put $E_{\mathrm BB} = E_{\mathrm n} = 0.21$ eV.
The temperature  was $T=375$ K, the 
diffusion prefactor $r_0=2 \times 10^{11} \mathrm{s}^{-1}$
($r_0=4 \times 10^{12} \mathrm{s}^{-1}$) for model I (model II) and the 
deposition flux was varied in the range \mbox{$F = 0.005 - 1.0$ ML/s.}
These choices were motivated mainly by our desire to 
study the variation of the meander wavelength over a range of control 
parameters without being strongly affected by finite size effects and
limited computer time. In particular, 
the different diffusion prefactors for the two 
models were chosen only to be able to study both models in the same
range of flux and temperature, and do not carry any physical
significance.  

Typical configurations generated in the simulations 
are shown in Figs.\ref{3d-prof_I},\ref{3d-prof_II} and \ref{averprof}.
For both models the initially straight steps form an in-phase meander pattern
with a characteristic wavelength. 
The selected wavelength remains constant during growth. 
The dependence of the meander
wavelength on the model parameters
is the main focus of the following discussion.
Some aspects of the temporal evolution of the pattern will be
addressed in Sect.\ref{Time}.

\section{Relevant length scales}
\label{Predictions}

Before turning to the quantitative analysis of the simulations, 
we summarize the available theoretical predictions
for the length scale of the meander instability. 

\subsection{The Bales-Zangwill instability}

The BZ analysis proceeds by solving
a diffusion equation for the adatom concentration with boundary conditions 
given by the attachment-detachment kinetics 
at the steps. It predicts an in-phase meander \cite{Pimpinelli94} 
with the dominant wavelength \cite{Gillet00}
\begin{equation}
 \label{linstab}
  \lambda_{BZ} = 4 \pi \sqrt{
   \frac{\Gamma \left(\Omega l D  
    c_{eq}^{0}+a \sigma_{st} \right) }{\Omega f_S F l^{2} }
 }.
\end{equation}
Here $\Omega$ denotes the atomic area and $a$ is the lattice spacing.
All kinetic and thermodynamic 
parameters entering (\ref{linstab}) can be expressed in terms
of the microscopic rates and energies of the KMC model.
The diffusion coefficient on the terrace reads $D = r_0 \exp(-E_{\mathrm S}/k_B T)$, 
and the step edge stiffness is given by\cite{Jeong99} 
$\Gamma= 2 a \sinh^{2}(E_{\mathrm K}/2k_BT)$.
To calculate the equilibrium adatom concentration $c_{eq}^0$ 
and the mobility $\sigma_{st}$ along the step edge, it
is useful to consider the step as a one-dimensional
(1D) SOS interface in equilibrium
with the adatom gas on the terrace. For model I the transition
rates are of Arrhenius form, i.e. dependent only on the configuration
before the jump,
and exact results for $c_{eq}^0$ and $\sigma_{st}$ can
be found \cite{Krug95}. We obtain
$c_{eq}^{0}=\exp(-2E_{\mathrm n}/k_BT)$ and
\mbox{$\sigma_{st} = (D/2) \exp(-2E_{\mathrm n}/k_B T)$}.

For model II the transition rates are no longer of Arrhenius type, and
the exact results \cite{Krug95} cannot be directly applied. 
Within linear fluctuation theory, the step edge 
mobility has been estimated as $\sigma_{st}=a^2/\tau_{L}$, where
$\tau_{L}$ is the 
characteristic time for detachment from a kink 
\cite{Pimpinelli93}.
\begin{figure}
\begin{center}
\includegraphics[width=4.5cm,height=8cm, angle = -90]{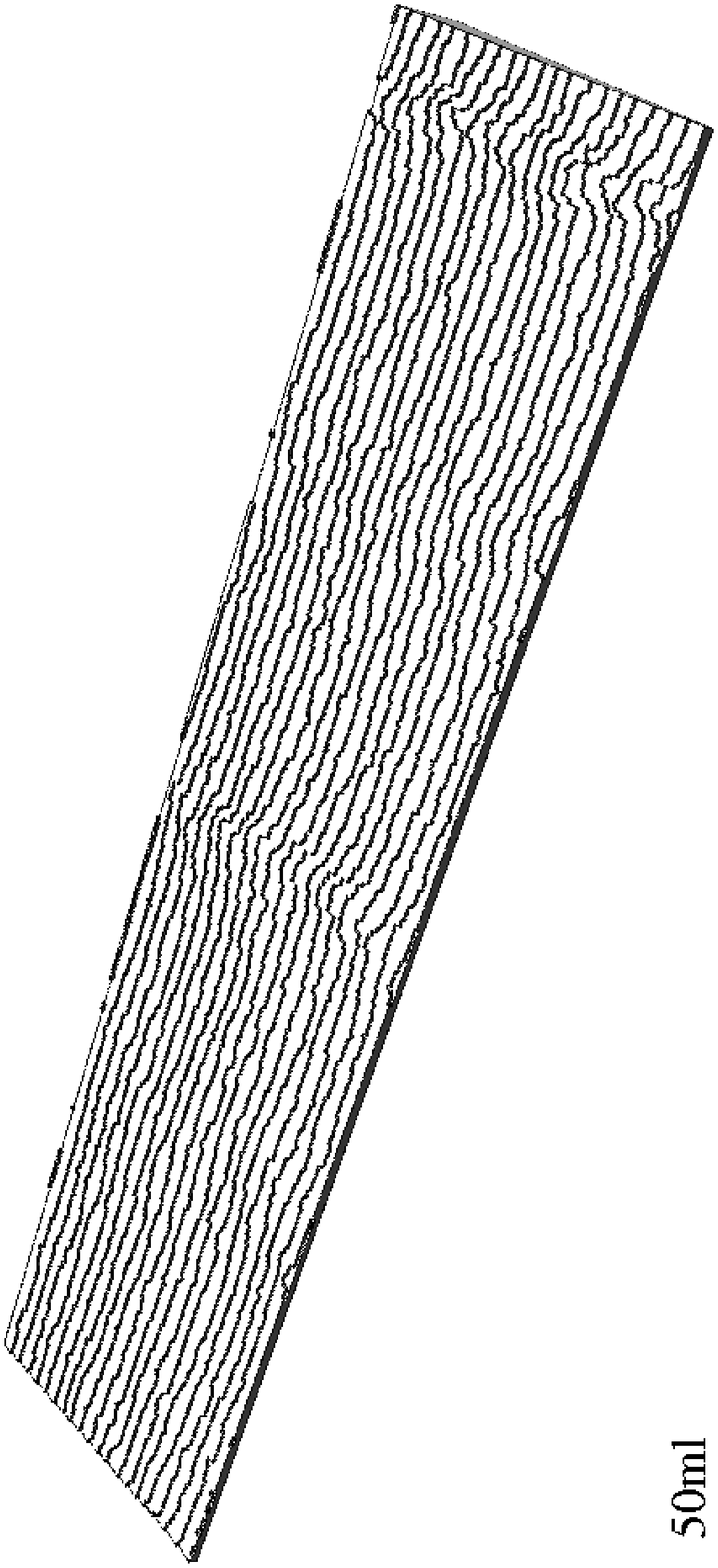}
\includegraphics[width=4.5cm,height=8cm, angle = -90]{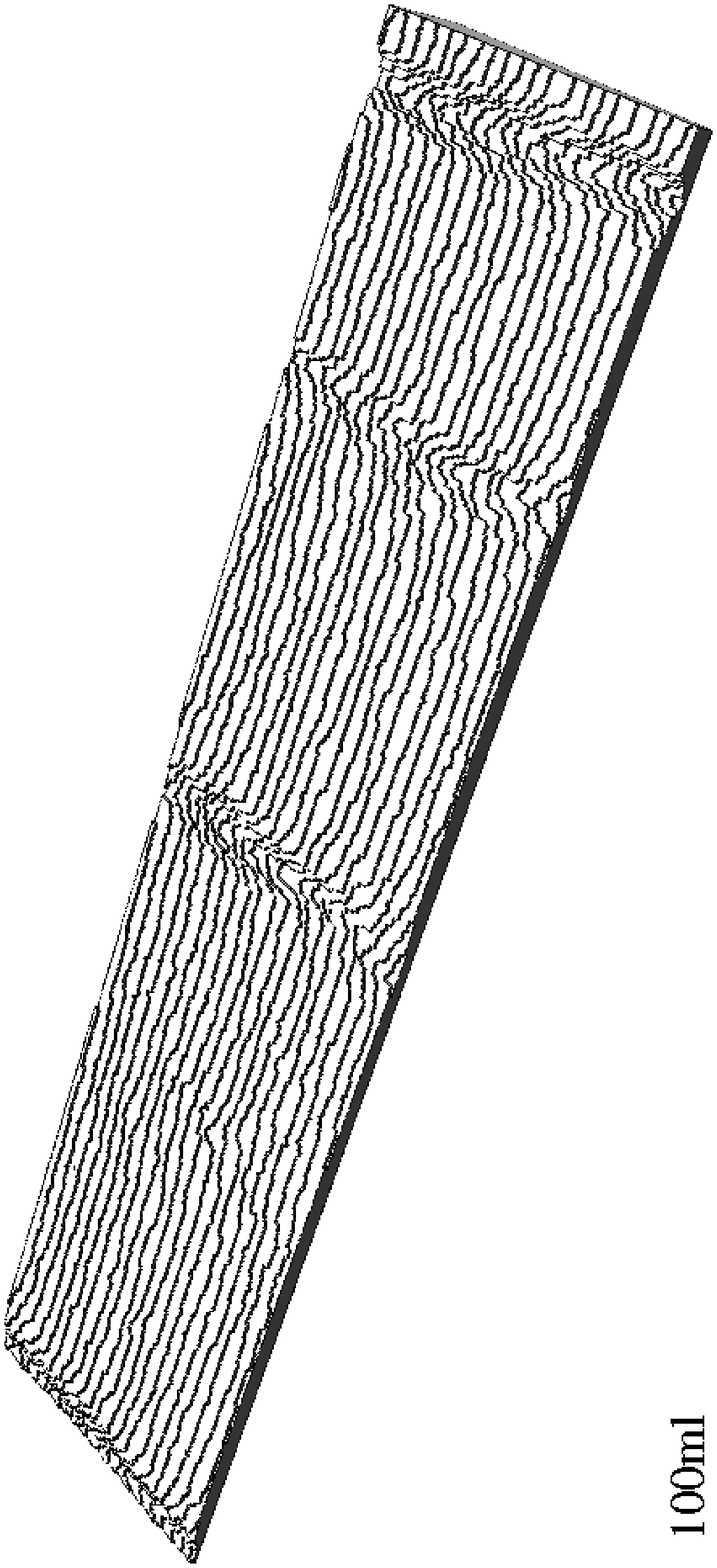}
\includegraphics[width=4.5cm,height=8cm, angle = -90]{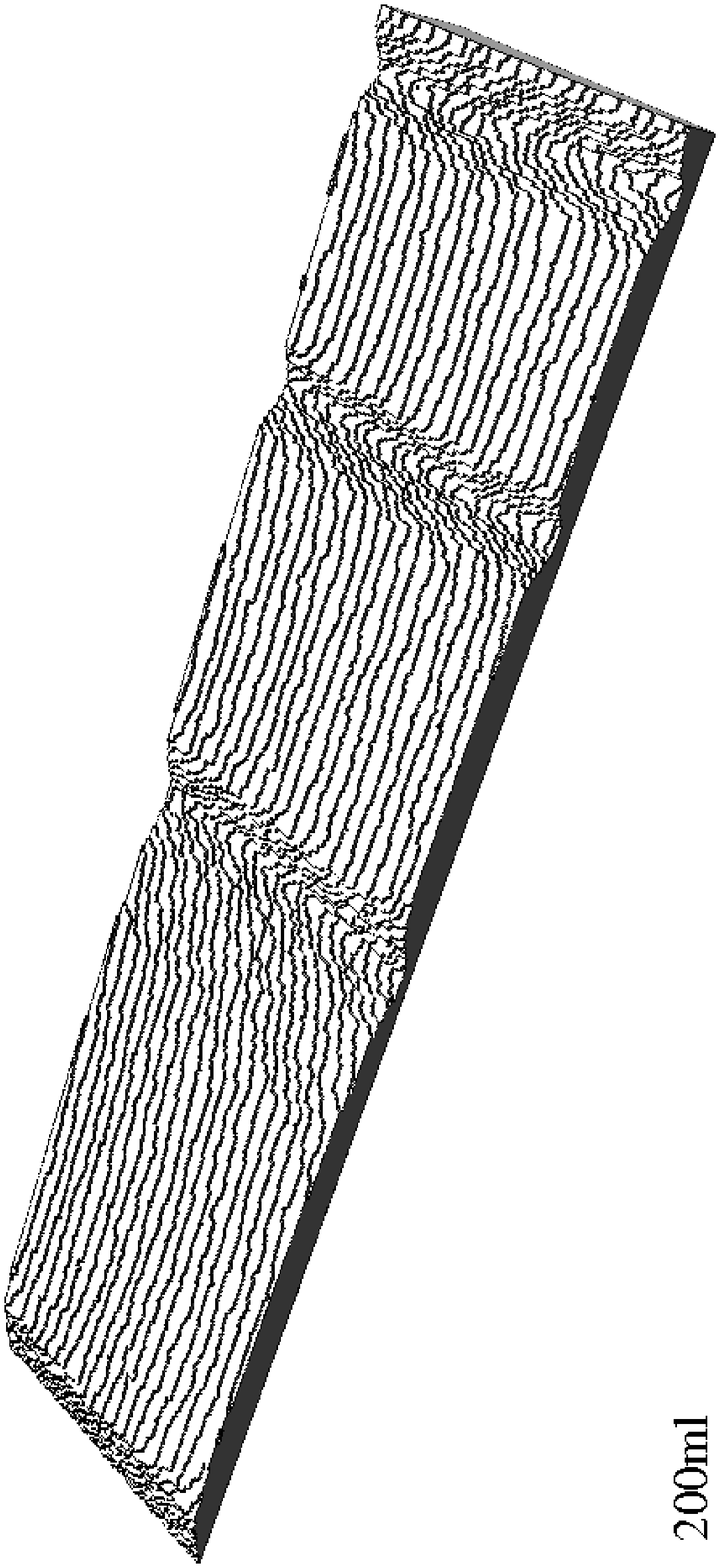}
\caption{Evolution of a vicinal surface with initially straight steps
under the dynamics of model I. Snapshots show
a $120 \times 500$ piece of a larger system ( $120 \times 1000$ with 20 steps) after
deposition of 50, 100 and 200 monolayers.
The deposition flux was $F=0.2$ monolayers per second (ML/s), 
and other parameters as described in the text.       
}
\label{3d-prof_I}
\end{center}
\end{figure}
\noindent
For our model II this yields
\begin{equation}
\label{mob}
\sigma_{st} \approx D \exp\left[ -(2E_{\mathrm n}+E_{\mathrm BB})/k_B T \right].
\end{equation}
Strictly speaking, 
(\ref{mob}) has to be modified in the presence of kink ES barriers
which reduce mass transport along the edge. A detailed
analysis\cite{KK01} shows that an additional
KESE barrier $E_{\mathrm KES}$
becomes relevant when $E_{\mathrm KES} > E_{\mathrm K}$. In our simulations
$E_{\mathrm KES} \approx E_{\mathrm BB} \leq E_{\mathrm K}$ always, 
so (\ref{mob}) suffices.
The terrace adatom concentration is generally
given by the expression $c_{eq}^{0}=\exp(-\Delta E/k_BT)$, where 
$\Delta E= 4 E_{\mathrm K}$ is the formation energy for moving an
adatom from a kink to the terrace \cite{Ghez88}.

The strength of the ES effect is contained in the parameter 
$f_S = (l_{-}-l_{+})/(l+l_{-}+l_{+})$,
where the length scales \cite{Krug97,Ghez88} $l_{\pm} =  D/k_{\pm}$ 
are inversely proportional to the 
\begin{figure}
\begin{center}
\includegraphics[width=4.5cm,height=8cm, angle = -90]{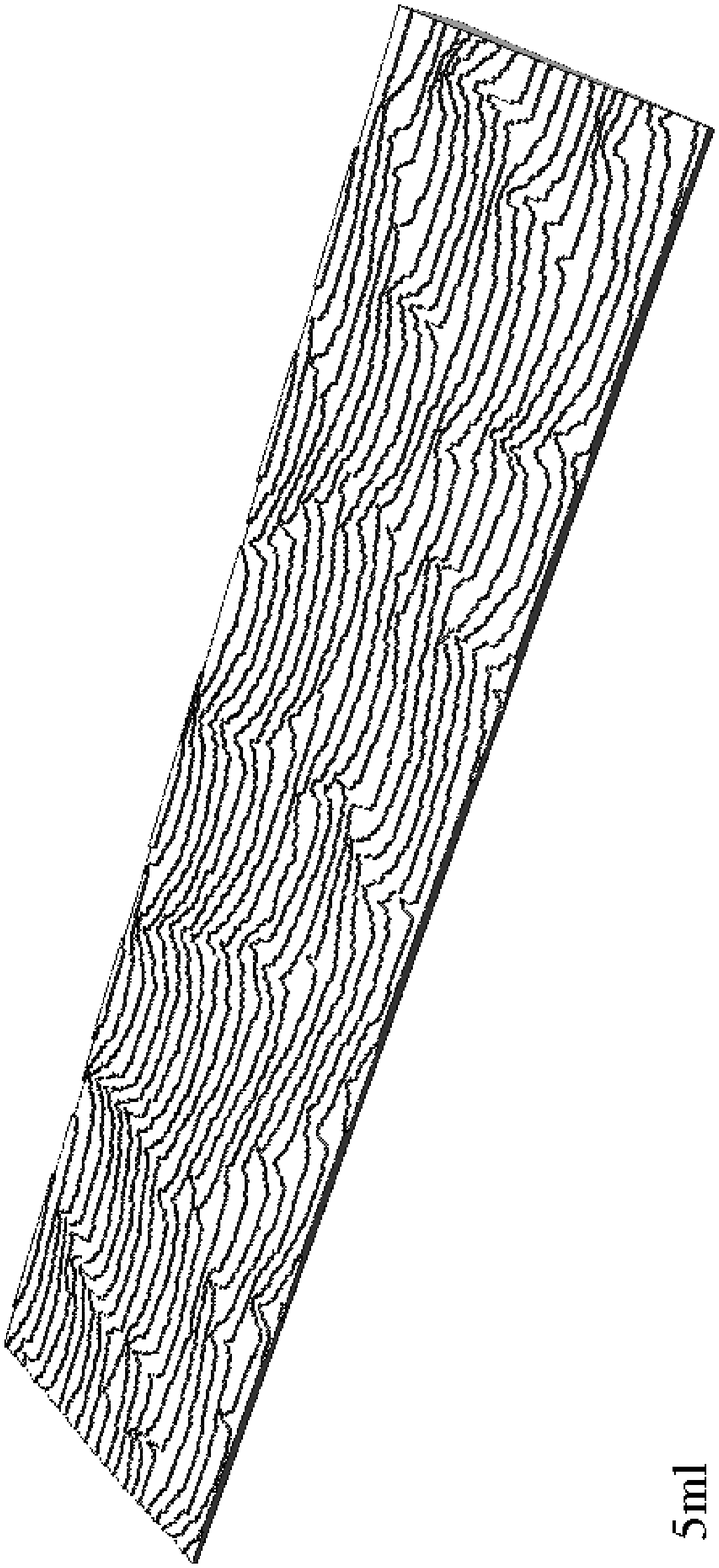}
\includegraphics[width=4.5cm,height=8cm, angle = -90]{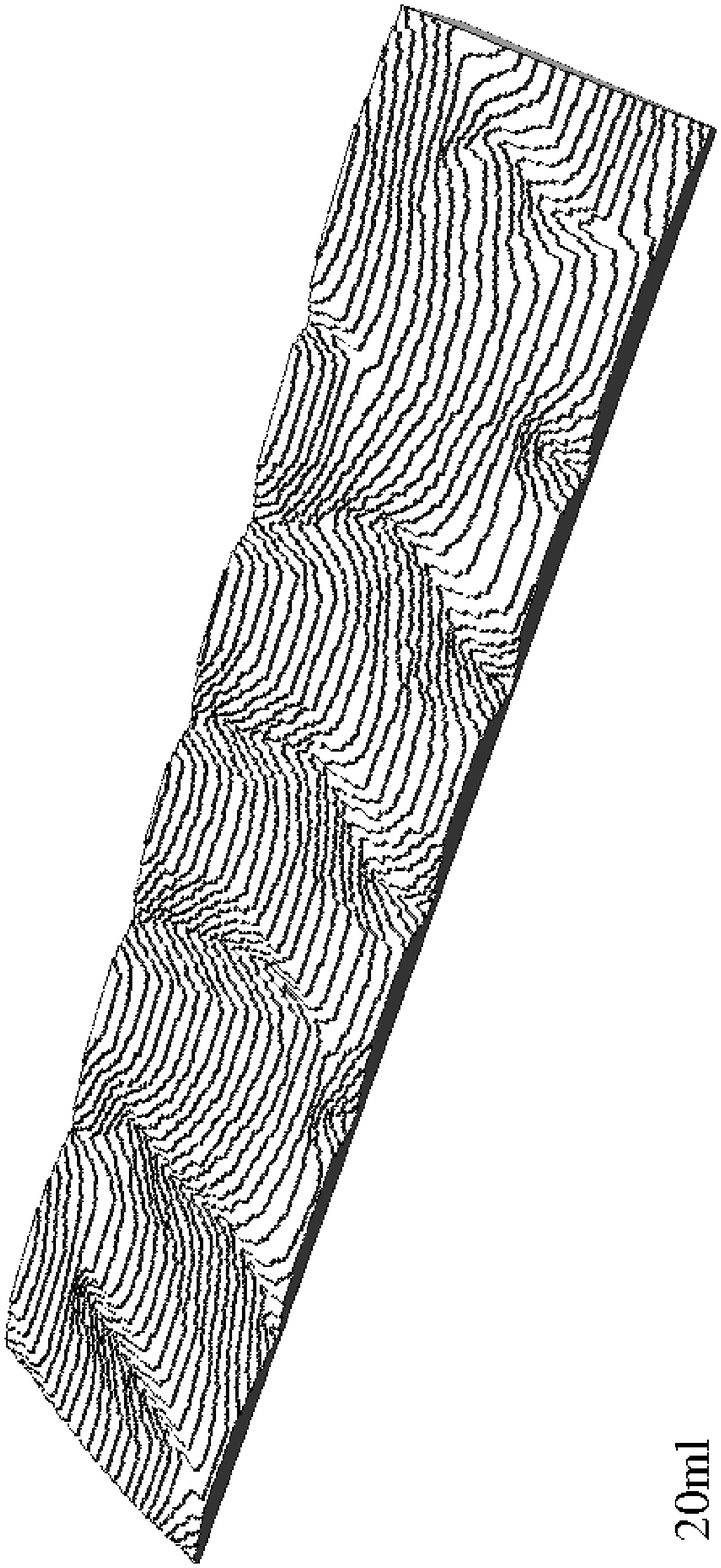}
\includegraphics[width=4.5cm,height=8cm, angle = -90]{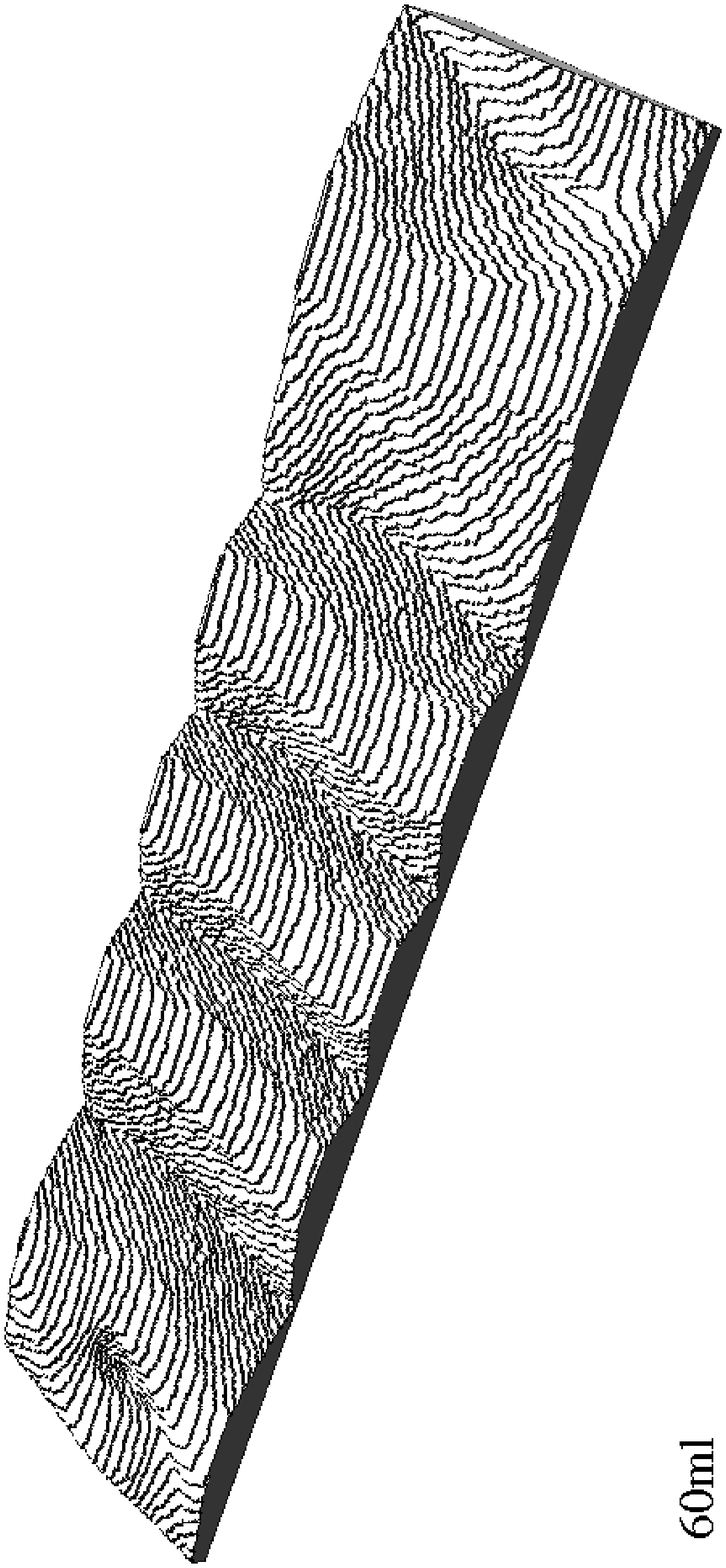}
\caption{ Evolution under the dynamics of model II. Snapshots show
a $120 \times 500$ piece of a larger system  
($120 \times 800$ with 20 steps) after
deposition of 5, 20 and 60 monolayers.
The deposition flux was $F=0.01$ ML/s
and other parameters as described in the text.    
}
\label{3d-prof_II}
\end{center}
\end{figure}
\noindent
attachment rates
for adatoms approaching the step from below ($k_+$) or
above ($k_-$)  
($l_{-}$ is also known \cite{Politi00b} as the ES length 
$l_{ES}$). For the KMC model we estimate
\begin{equation}
\label{k-coef}
k_{+}=D \ , \ k_{-} \approx D \exp(-E_{\mathrm ES}/k_BT). 
\end{equation}
In fact the rate of attachment from the upper terrace depends on the
microscopic configuration of the step. A direct test of the relation
(\ref{k-coef}) is presented below in Sect.\ref{Schwoebel}.   

The expression used above for the edge mobility
$\sigma_{st}$ is valid only on length scales longer than the kink
distance $l_{K} \approx (1/2) \exp(E_{K}/k_{B} T)$. 
On shorter scales edge diffusion is much more efficient,
and the expression for the characteristic meander wavelength 
must be replaced by\cite{Pierre-Louis01} 
\begin{equation}
 \label{OPL-edge-diff}
  \tilde{\lambda}_{BZ} = 2^{1/4} \ \sqrt{l_K \pi} \ \lambda_{BZ}^{1/2} 
\end{equation}     
when edge diffusion dominates and
$l_K \gg \lambda_{BZ}$.

\subsection{The KESE instability}

Step meandering due to the KESE can be discussed \cite{Pierre-Louis99}
in analogy to the ES instability of a one-dimensional surface 
\cite{Politi96,Politi00b}. 
The characteristic
wavelength depends on the one-dimensional \emph{nucleation length}
$l_D$, which is defined as the average distance between two dimers that
are nucleated on a straight step at the beginning of deposition,
and the kink ES length $l_{\mathrm{KES}} \approx \exp[E_{\mathrm{KES}}/
k_B T]$. For a strong KESE, in the sense that $l_D \ll l_{\mathrm{KES}}$,
the initial meander wavelength is $l_D$, while for a weak KESE 
($l_D \gg l_{\mathrm{KES}}$) 
it is of the order of \cite{Politi96,Krug97} 
\begin{equation}
\label{weak1}
\lambda_w \approx (\ell_D/\ell_{\mathrm{KES}})^{1/2} \ell_D.
\end{equation}
From one-dimensional nucleation theory  
the expression
\begin{equation}    
 \label{nuclen}
  l_D = (12 D_{st}/Fl)^{1/4}   
\end{equation}
for the nucleation length can be derived \cite{Politi97}; here
$F l$ is the flux of terrace atoms onto the step. 
In contrast to the mobility $\sigma_{st}$, the
edge diffusion coefficient $D_{st}$ refers to the motion of an 
edge atom along a \emph{straight} step without kinks, and is given
by $D_{st} =  D \exp(-E_{\mathrm n}/k_B T)$ in the SOS model. 

It should be noted that (\ref{weak1},\ref{nuclen}) apply to a
1D surface in the absence of desorption; for a step this translates into 
neglecting the detachment from the step. This approximation clearly breaks 
down as the bond breaking barrier $E_{\mathrm BB}$ approaches zero. 
Including the detachment from the step introduces a new length scale 
into the problem, namely the diffusion length 
$x_s=\sqrt{D_{st} \tau}$. Here $\tau$ is the time an adatom
diffuses along a straight step before being ``evaporated'' to the
terrace, and $x_s$ the distance the atom travels along the step in
time interval $\tau$. A lower bound for $\tau$
is given by the detachment rate from the straight step 
$\tau^{-1} = D_{st} \exp(-E_{\mathrm BB}/k_B T)$. In reality detached atoms
have a high probability of re-attachment and thus the real 
evaporation time $\tau$ is longer. A lower bound for the diffusion length 
then reads $x_{s} = \exp(E_{\mathrm BB}/2 k _B T)$.  
Repeating the calculation of the nucleation length for a 1D surface with 
desorption \cite{KK01}, 
one arrives at an expression for the nucleation length
in the limit $x_s \ll l_D$, which reads 
\begin{equation}    
 \label{nuclen_det}
  l_D = \left(\frac{D_{st}}{Fl} \right)^{1/2} \frac{1}{x_s} ,  
\end{equation} 
The same result can be obtained also by applying the scaling arguments 
of Jensen \emph{et al.} \cite{Jensen97} in one dimension.
For general values of $x_s$, $l_D$ is given as the solution of 
a transcendental equation. 
A simple expression interpolating between the two cases (\ref{nuclen})
and (\ref{nuclen_det}) reads
\begin{equation}    
 \label{nuclen_intrp}
  l_D = \left(\frac{D_{st}}{Fl} \right)^{1/4}
  \left[12^{1/4} +  \frac{1}{x_s} 
   \left( \frac{D_{st}}{Fl}\right)^{1/4} \right].  
\end{equation} 
In the case of a weak KESE, repeating the calculation
of Politi and Villain \cite{Politi96} with desorption, one
finds \cite{KK01} that for 
\begin{figure}[t!]
\begin{center}
\includegraphics[width=6.5cm,height=7.5cm, angle = -90]{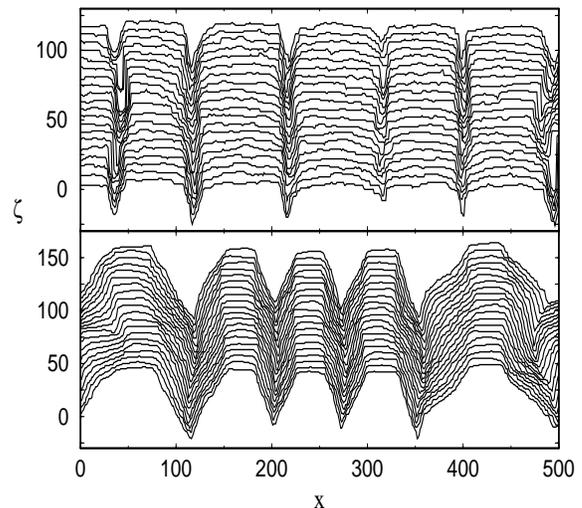}
\caption{Step configurations for model I (upper panel) and model 
II (lower panel) after deposition of 75 ML at a 
deposition flux $F=0.5$ ML/s (model I) and 
$F=0.01$ ML/s (model II). The figures show part of a
800 $\times$ 120 lattice with 20 steps.}
\label{averprof}
\end{center}

\end{figure}
\noindent
$x_s \ll l_{\mathrm KES} \ll l_D$
the most unstable wavelength is of the order
\begin{equation}
\label{weak}
\lambda_w \approx l_D^2/ x_{s},
\end{equation}
which replaces (\ref{weak1}). 

\section{Simulation results}
\label{Results}
\subsection{Meander mechanisms}

\begin{figure}
\begin{center}
\includegraphics[width=5.3cm,height=7.5cm, angle = -90]{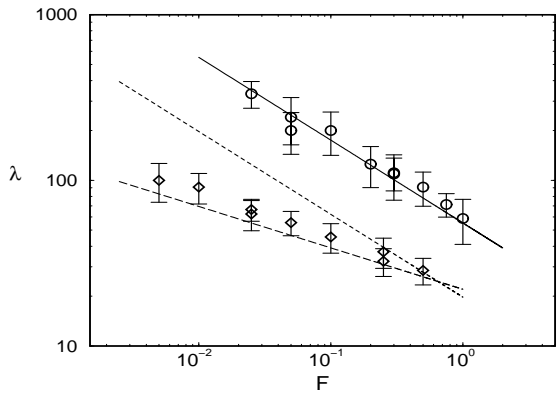}
\caption{Meander wavelength as a function of flux for
model I (circles) and model II (diamonds).  
Each symbol represents a single run on a lattice of 
size 1000 $\times$ 30 with 5 steps. The error bars refer to 
the variation of the wavelength within the configuration. For some fluxes
results for a lattice of size 30 $\times$ 1200 have been included also. 
The simulations were run until the meander wavelength was clearly 
visible. The BZ-length (\ref{linstab}) is plotted as a 
full line for model I and  a short-dashed line for model II.
The long-dashed line is the nucleation length 
(\ref{nuclen}) for model II.}
\label{wavelen}
\end{center}
\end{figure}

Figure \ref{wavelen} shows the meander wavelength as a function of deposition 
flux obtained from our KMC simulations.
The wavelength was extracted from the profiles directly by measuring the distance
between subsequent minima in a single profile. The error bars refer to 
the variation of the wavelength within a single profile.
For model I the wavelength is found to scale as $F^{-0.47 \pm 0.06}$,
in qualitative agreement with both the BZ-length $\lambda_{BZ}$ and 
the nucleation length $l_{D}$ 
in the detachment-dominated limit 
(Eq.(\ref{nuclen_det}); for model I $x_{s}\approx 1 \ll l_{D}$).
Quantitatively the results are found to agree with $\lambda_{BZ}$
for the parameters used in the simulations, while 
the nucleation length $l_{D}$
is smaller by approximately a factor $1/2$. A more convincing piece of
evidence for the Bales-Zangwill mechanism is the dependence of the 
meander wavelength on the Ehrlich-Schwoebel barrier $E_{\mathrm ES}$, which is 
discussed below in 
Section \ref{Schwoebel}. 
The nucleation length is obviously independent of $E_{\mathrm ES}$.       

For model II the meander wavelength scales as $F^{-0.28 \pm 0.05}$, 
which disagrees with the BZ theory but is
\noindent
consistent both with the
modified BZ length (\ref{OPL-edge-diff}), and with the nucleation length
(\ref{nuclen}) in the absence of detachment. 
However, Eq.(\ref{OPL-edge-diff}) predicts a prefactor
that is one order of magnitude too large. 
This is not surprising, since (\ref{OPL-edge-diff}) was derived for steps
close to thermal equilibrium \cite{Pierre-Louis01}; 
under growth conditions the kink density is 
much larger than its equilibrium value.

Including detachment and using the lower bound  
$x_{s} = \exp(E_{\mathrm BB}/2 k _B T)$ 
as an approximation for the diffusion length yields an upper bound for 
the nucleation length. Expressions (\ref{nuclen}) and (\ref{nuclen_intrp})
thus give lower and upper bounds for the nucleation length, which differ
approximately by a factor of 2. The lower bound (\ref{nuclen}) 
is seen to quantitatively describe the simulation data for model II,
which shows that for model II the detachment may be neglected.  
This confirms that one-dimensional nucleation is the
relevant wavelength selection mechanism under conditions of facile 
edge diffusion, in accordance with the conclusions from 
previous experimental \cite{Maroutian99}
and simulational \cite{Rusanen01,Rusanen02} work on surfaces vicinal to 
Cu(100).

The meander wavelength by itself
does however not uniquely specify the instability mechanism.
The BZ theory predicts a band of unstable wavelengths
extending over the interval $(\lambda_{min}, \infty)$,
where $\lambda_{min} = \lambda_{BZ}/\sqrt{2}$ and  
$\lambda_{BZ}$, as given by (\ref{linstab}), 
is the wavelength of perturbations with the
maximal growth rate. Numerical studies of a nonlinear evolution equation
for the in-phase meander show that initial wavelengths
between $\lambda_{min}$ 
and $\lambda_{max} \approx 3 \lambda_{BZ}$ are
preserved during growth \cite{Kallunki00}. Thus deviations from the 
BZ prediction (\ref{linstab}) can be attributed partly to 
a wavelength different from $\lambda_{BZ}$ which dominates
the spectrum of initial perturbations. In this context it is important
to note that, for small fluxes, the 
\begin{figure}
\begin{center}
\includegraphics[width=5.3cm,height=7.5cm, angle = -90]{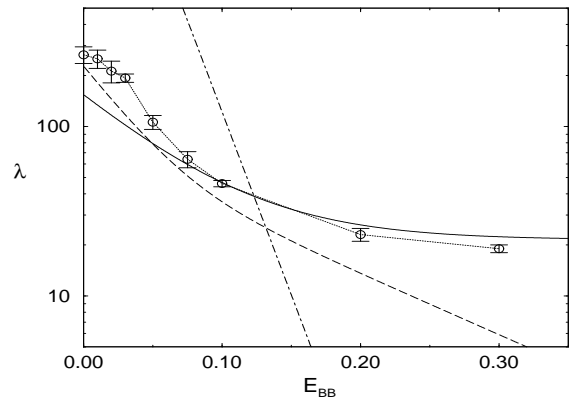}
\caption{The observed wavelengths (circles)  for various values of the bond breaking
barrier $E_{\mathrm BB}$. The full line is the nucleation length (\ref{nuclen_intrp}), the
dashed line the BZ-length (\ref{linstab}) and the dot-dashed line the meander 
wavelength (\ref{weak}) for weak KESE. The crossover between the two 
meander mechanisms occurs around $E_{\mathrm BB} \approx 0.05$ eV. 
Each point is an average over 5 
independent runs on a $250..1500 \times 30$ lattice with 3 steps 
(step spacing $l=10$).   
}
\label{wl_vs_Ebb}
\end{center}
\end{figure}
\noindent
wavelength measured for 
model II is \emph{smaller} than the minimal unstable wavelength
$\lambda_{min}$ of BZ theory; this can be seen by shifting the
short-dashed line in Fig.\ref{wavelen}, which represents $\lambda_{BZ}$
for model II, downward by a factor of $\sqrt{2}$. 
This proves that for model II an instability mechanism
different from the BZ mechanism -- the KESE -- is the cause of the meander.

\subsection{Crossover between the two mechanisms}

Which of the two meander mechanisms is operative depends on the
importance of step edge diffusion and on the strength of the KESE barrier. 
In our SOS model both are controlled by the bond breaking barrier
$E_{\mathrm{BB}}$. Thus decreasing 
$E_{\mathrm{BB}}$ should lead to a crossover from
the KESE instability to the BZ instability. 

Simulation results for various values of 
$E_{\mathrm BB}$ are shown in Fig. \ref{wl_vs_Ebb}.
The wavelength was determined by counting the number
of minima in the step profile on a finite sample.  
Reported wavelengths are averages over 5 independent runs and
the error bars are the standard deviations.
Since the bond breaking barrier enters the kink energy through (\ref{EK}), 
the predicted length scales (\ref{linstab}) and (\ref{nuclen_intrp})
both increase with decreasing $E_{\mathrm{BB}}$ in a qualitatively
similar manner. In addition, the decrease of the KESE barrier
implies a transition from the strong KESE to the weak KESE regime.
The numerical data clearly show that 
for $E_{\mathrm BB}>0.10 \, \mathrm{eV}$ the meander wavelength is set by
the nucleation length $l_{D}$, while for $E_{\mathrm BB} < 0.05$ eV  
the simulations are consistent only with the BZ length, because the 
nucleation length is considerably smaller, and 
\begin{figure}
\begin{center}
\includegraphics[width=5.3cm,height=7.5cm, angle = -90]{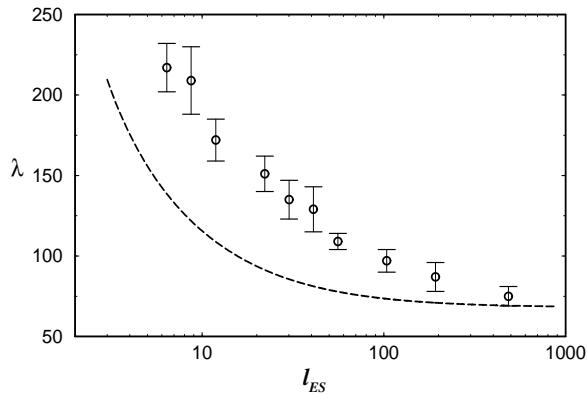}
\caption{Meander wavelength as a function of the
ES length $l_{ES} = D/k_-$ for model I at $F = 0.25$ ML/s. 
Each symbol is an average over 5 runs on a lattice of size
$1250 \times 15$ containing a single step
(step spacing $l=15$). The dashed line shows
the BZ prediction (\ref{linstab}).  
}
\label{finite_les}
\end{center}
\end{figure}
\noindent
the weak KESE length (\ref{weak})
is much larger than the actual meander wavelength.

\subsection{Variation of the ES barrier
\label{Schwoebel}}
As a further test of the BZ prediction (\ref{linstab}) we have
measured the meander wavelength in model I for different values
of the ES barrier.
At $T=375$ K and with the barrier $E_{\mathrm{ES}} = 0.15$ eV used above,
the Schwoebel length $l_{-} \approx 104 \gg l$ and $l_{+}=1$,
so that $f_S \approx 1$ in Eq. (\ref{linstab}). To study
the effect of a finite ES length we carried out simulations 
with a single terrace\cite{Single_note} of width $l =15$, varying the ES barrier between 
$E_{\mathrm{ES}} = 0.06$ eV and  0.20 eV ($l_{-} = 6  - 487 $), 
while keeping the other barriers at the values given previously.
Again, the average wavelengths were calculated from 5 independent runs.
Figure \ref{finite_les} shows that the dependence of the
meander wavelength on the ES barrier is qualitatively described by BZ theory,
but Eq.(\ref{linstab}) is not quantitatively accurate. The 
true ES length appears to be smaller than that given in (\ref{k-coef}) by
about a factor 1/4, which corresponds to a reduction of the ES barrier by
0.04 eV. This cannot be a simple effect of step roughness, since 
the implementation of the ES barrier used in the present work in fact
implies that the approach to a kinked step involves a \emph{higher}
barrier than to a straight step. This issue deserves further consideration.

\subsection{Temporal evolution}
\label{Time}

The two models can be distinguished
with regard to the dynamics of meander formation.
For model I the steps meander in-phase from the beginning, whereas
for model II the meander starts with random phase shifts between the
steps (compare the early time configurations
in
\begin{figure}
\begin{center}
\includegraphics[width=5.5cm,height=7.5cm, angle = -90]{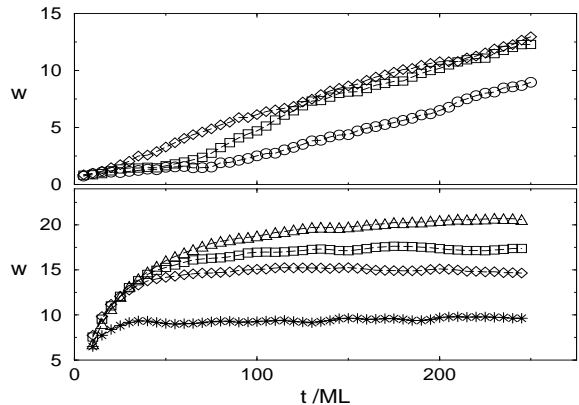}
\caption{ Average step width $w$ as a function of time for model I
(upper panel, $F = 0.1 ( \bigcirc ), 0.4 ( \square ), 0.8 (\diamond )$ ML/s)
and model II (lower panel, $F=0.05 ( \star ),0.2 ( \diamond ),
0.4 ( \square ) ,0.8 ( \triangle ) $ ML/s), from single runs on a lattice
of size $500 \times 30$ with 5 steps.
}
\label{wi_vs_ml}
\end{center}
\end{figure}
\noindent
Figures \ref{3d-prof_I} and \ref{3d-prof_II} ). 
Later on, as the meander amplitude grows and the step-step interaction
through the diffusion field becomes effective, the correlations between the steps
grow, leading asymptotically to an in-phase step train. In the light of the two
different mechanisms these observations are easy to understand:
Since for model I the meander is due to the 
BZ instability, the step train is expected
to be in-phase from the beginning, this being the fastest growing mode 
\cite{Pimpinelli94}; for model II the meander starts independently at each step since
the meander is due to the local adatom dynamics at the step edge. 

We turn next to the time evolution of the meander amplitude, which
also differs for the two models (Fig.\ref{wi_vs_ml}).
We consider the step width $w$ defined by
\begin{equation}
  \label{defwidth}
   w^2 \equiv L^{-1} \sum\limits_{x=1}^{L} \left[ y(x) \right]^{2} ,
\end{equation}
where the $x$-coordinate is directed along the step, $L$ is the step length
and $y(x)$ is the step position relative to its mean.
For model II, $w(t)$ increases very rapidly at the beginning,
but asymptotically saturates \cite{Note:Rusanen}. 
In contrast, for model I 
the step width was found to increase linearly in time 
for the longest times we could access\cite{longtime-note}.
Both observations are at variance with the predictions  
of a nonlinear step evolution equation for the in-phase
meander \cite{Pierre-Louis98b,Kallunki00,Gillet00},
which predicts that $w \sim \sqrt{t}$. 

Our results also contradict earlier Monte-Carlo simulation
\cite{Pierre-Louis98b}, in which
the steps were described by single valued functions, thus
prohibiting step overhangs and voids, and nucleation
on the terraces was not allowed for. In these simulations
a regime with $w \sim \sqrt{t}$ was observed at long times.
A direct comparison of the two sets of simulations is not 
possible, however, because of the rather different choice 
of parameters.  
In the earlier work \cite{Pierre-Louis98b},   
the concentration of adatoms was set to a value $c_{eq}^{0}=0.119$,
which is several orders of magnitude larger than in our simulations.
In fact, at such high adatom concentrations step flow is hardly possible,
if the nucleation on the terraces is not artificially suppressed. 
For this reason we cannot reproduce the conditions used in the
earlier simulations\cite{Pierre-Louis98b}. 

The linear amplitude growth in model I suggests that minima and
maxima of the step meander move at different velocities.
This may be due to diffusional screening which prevents adatoms 
to reach the narrow fjords separating the cells in the upper panel of
Fig.\ref{averprof}. In model II the fjords are wider, because adatoms
are able to fill them up through edge diffusion. 
The saturation of the amplitude in model II 
may be related to the stabilizing character of the KESE step current
at large slopes \cite{Pierre-Louis99,Politi00a}.

\section{Relation to experiments on Cu(100)}
\label{Exps}

In this section we briefly comment on the relevance of our work
for the experimentally observed meander instability 
\cite{Maroutian01} on surfaces vicinal to Cu(100).
In these experiments two different vicinal surfaces were considered,
which consist of dense packed $\langle 110 \rangle$ steps, and 
open $\langle 100 \rangle$ steps, respectively.

For the dense packed $\langle 110 \rangle$ steps, edge diffusion is much more facile
than detachment, so the scenario of our model II should apply.
Indeed, using the expression (\ref{nuclen}) for the one-dimensional
nucleation length to interpret the experimentally measured activation
energy $E_a = \, 0.092 \, \mathrm{eV}$ of the meander wavelength,
$\lambda \sim e^{-E_a/k_B T}$, 
one obtains an energy barrier $E_{st} = 4 E_a = 0.37 \, \mathrm{eV}$ for 
diffusion along a straight edge, which is consistent with the
estimate $E_{st} = 0.45 \pm 0.08 \, \mathrm{eV}$ derived from the
analysis of time-dependent STM observations \cite{Giesen94}. 
Also the flux dependence of (\ref{nuclen}) as $F^{-1/4}$  
agrees with the experimental power law exponent of $- 0.21 \pm 0.08$.  

What is missing to complete the picture is some direct experimental
evidence for a (strong) KESE 
at the $\langle 110 \rangle$ step \cite{Note:Rh}. 
Here we want to point
out that indirect evidence for a kink ES barrier follows from
a comparison of the growth experiments \cite{Maroutian01}
with step fluctuation measurements. Using the
accepted value \cite{Giesen01} $E_{\mathrm K} \approx 0.13 \, \mathrm{eV}$ for
the kink energy, the measurement of the prefactor of the temporal
step correlation function \cite{Giesen95} yields the estimate
$E_\sigma \approx 0.91 \, \mathrm{eV}$ for the activation energy 
of the step edge mobility $\sigma_{st}$. It was mentioned above
in Section \ref{Predictions} that, in the absence of a strong
KESE, this can be identified with the energy barrier $E_{\mathrm{det}}$
for the detachment of a step adatom from a kink. In a simple bond counting 
approximation (which is supported by effective medium (EMT) 
\cite{Merikoski97} and embedded atom (EAM) \cite{Mehl99} calculations)
the detachment barrier is given by $E_{\mathrm{det}} \approx 
E_{st} + 2 E_{\mathrm{K}}$. Using the value
of $E_{st}$ determined from the meander wavelength of the $\langle 110 \rangle$ step
this yields $E_{\mathrm{det}} \approx 0.63 \, \mathrm{eV}$, which is much
smaller than the step fluctuation estimate of $E_\sigma$. 

The discrepancy
strongly suggests that the migration of atoms along the kinked step
is suppressed by an additional kink ES barrier $E_{\mathrm{KES}}$. 
The quantitative analysis \cite{KK01} shows that for 
$E_{\mathrm{KES}} > E_{\mathrm{K}}$, the activation energy for 
$\sigma_{st}$ is given by $E_\sigma = E_{\mathrm{det}} + E_{\mathrm{KES}}-
E_{\mathrm{K}}$, which, using the numbers quoted above, yields
the estimate $E_{\mathrm{KES}} \approx 0.41 \, \mathrm{eV}$. In agreement
with semiempirical calculations \cite{Merikoski97,Mehl99}, 
the additional kink barrier is
found to be comparable to the barrier $E_{st}$
for diffusion along a straight step.   

The situation is rather different 
for the open $\langle 100 \rangle$ step, at which a meander instability
with similar characteristics has been observed \cite{Maroutian01}.
Maroutian \emph{et al.}  
identify the measured wavelength for the open step with 
the one-dimensional nucleation length (\ref{nuclen}).
This interpretation seems questionable for the following
two reasons. First, 
since the open step can be viewed as being composed entirely of 
$\langle 110 \rangle$ kinks,
the edge diffusion barrier $E_{st}$ is expected to be 
much larger than along the 
close packed step \cite{Trushin97}, while the additional barrier for 
detachment from a kink (which is related to the energy of $\langle 100
\rangle$ kinks) is much smaller \cite{Mehl99}. Under these conditions
the initial growth of the step is not well described by the picture of 
one-dimensional nucleation. Second, symmetry arguments
\cite{Pierre-Louis99,Politi00a} show that the step edge current
induced by the KESE should \emph{stabilize}, rather than destabilize
the open step. Thus the mechanism which could lead to an instability
on the scale of the nucleation length is absent.   

This leaves the BZ instability as the most
plausible instability mechanism \cite{Maroutian01b}. However, neither
the activation energy (as estimated from known energetic parameters
\cite{Giesen01,Merikoski97,Mehl99}) 
nor the $F^{-1/2}$-dependence predicted by 
(\ref{linstab}) matches the experimentally determined meander 
wavelength \cite{Maroutian99,Maroutian01}. 
In addition, in KMC simulations 
of Cu(100), the surface composed of $\langle 100 \rangle$ steps 
was found to be stable \cite{Rusanen01}; possibly this reflects
the stabilizing KESE. The experimentally observed
instability of the open step thus poses a major puzzle at the moment.

\section{Conclusion}
\label{Conclusions}

In summary, we have studied the unstable growth dynamics of vicinal surfaces
using an SOS model with two different choices of atomic processes. It 
was shown that in both cases the steps form an in-phase wave pattern. 
The formation mechanism and the
wavelength of the pattern depends on whether the 
diffusion of adatoms along the step edges is significant or not.  
In the case where step edge diffusion is negligible the wavelength was 
found to be correctly predicted by the 
continuum theory of BZ, while in the case of facile edge diffusion
(combined with an ES barrier at kinks) it is set by the one-dimensional
nucleation length. 

In both cases 
good agreement between the KMC simulations
and the analytic predictions (\ref{linstab}) and (\ref{nuclen})
for the meander wavelength was achieved \emph{without any adjustable
parameters}. However, the asymptotic time evolution of the
step profile disagrees with the prediction of an effective step 
evolution equation \cite{Pierre-Louis98b,Kallunki00,Gillet00}. Whether 
this disagreement is due to different physics described by the 
continuum and discrete models or due to the approximations made 
deriving the step equation remains to be clarified.

\begin{acknowledgments}
We are grateful to O. Biham, H.-J. Ernst,
J.W. Evans, M. Giesen, O. Pierre-Louis and
M. Rusanen for useful discussions, and to T. Maroutian
for sending us a copy of his thesis \cite{Maroutian01b}. 
J. Kallunki acknowledges the kind
hospitality of Universit\'e Joseph Fourier, Grenoble. 
This work was supported
by Volkswagenstiftung, by DFG within SFB 237, and by the COST project
P3.130. The final version of the paper was prepared at 
MPI f\"ur Physik komplexer Systeme, Dresden.
\end{acknowledgments}


\begin{references}
\bibitem{Single_note} Because of the periodic boundary conditions,
a single terrace is equivalent to an infinite step train with
steps that are perfectly in phase.
\bibitem{longtime-note} At times longer than those depicted in 
Fig.\ref{wi_vs_ml}, step flow
breaks down due to the formation of step loops and craters, and a transition to mound growth
sets in \cite{Rost96}. 
The details of this process will be addressed elsewhere.
\bibitem{Note:Rusanen} In a recent KMC simulation \cite{Rusanen02} 
of Cu(100), where step meandering has also been attributed
to the KESE, $w \sim t^{0.3}$ was observed at early times ($t < 5$ ML).
In earlier SOS model simulations \cite{Pierre-Louis99} the behavior
$w \sim t^{0.3}$ and $w \sim t^{0.57}$ was reported for weak and strong
KESE, respectively.
\end{references}
\end{document}